\documentclass[prl,twocolumn,superscriptaddress,preprintnumbers,
               amsmath,amssymb]{revtex4}
\usepackage{graphicx}
\usepackage{epsfig}
\usepackage{color}
\newcommand{\ea}{\mbox{\em et al. \/}}

\begin{document}

\title{On electrons and hydrogen-bond connectivity in liquid water}

\author{M. V. Fern\'andez-Serra}
\affiliation{Laboratoire de Physique de la Mati\`ere Condens\'ee 
             et Nanostructures (LPMCN), Universit\'e Claude Bernard Lyon 1,
             69622 Villeurbanne, France}\email{marivi@lpmcn.univ-lyon1.fr}
\author{Emilio Artacho}
\affiliation{Donostia International Physics Center,
             Universidad del Pa\'{\i}s Vasco,
	     20080 San Sebastian, Spain,}
\affiliation{Department of Earth Sciences,
             University of Cambridge,
             Downing street, Cambridge CB2 3EQ, UK}
\date{\today}

\begin{abstract}
  The network connectivity in liquid water is revised in terms of 
electronic signatures of hydrogen bonds (HBs) instead of 
geometric criteria, in view of recent X-ray absorption studies.
  The analysis is based on ab initio molecular-dynamics 
simulations at ambient conditions.
  Even if instantaneous thread-like structures are observed in the
electronic network, they continuously reshape in oscillations reminiscent 
of the $r$ and $t$ modes in ice ({$\tau$}$\sim$170 fs).
  However, two water molecules initially joint by a HB remain effectively 
bound over many periods regardless of its electronic signature.

\end{abstract}

\pacs{XX.XX.xx}

\maketitle


  Water is an extremely intriguing liquid that continues to
excite the interest of scientists in many disciplines. 
  Many of its anomalous properties \cite{Errington02,Speedy76}
originate in the hydrogen bonds (HBs) among water molecules 
\cite{stillingerwater,tip5p}. 
  The concept of a network liquid emerges naturally from this
HB connectivity, an intuitive image that has provided interesting 
insights into the properties of water \cite{Errington02, Luzar00}.
  Direct structural information to characterize such network structure 
is hard to obtain experimentally.
  Diffraction techniques \cite{waterexp2,waterexp1} offer radial 
distribution functions (RDFs) very naturally, but rely on reverse Monte Carlo
techniques using force-field models to obtain further structural information
\cite{Soper96}.
  Spectroscopic probes provide a rich source of complementary
information. 
  X-ray emission \cite{Guo02,Kashtanov04} (XES) and X-ray absorption 
\cite{Myneni02,Cavalleri02,Wernet04} (XAS) spectroscopies explore the 
electronic states of the liquid right below and above the Fermi level,
respectively.
  In particular, the work by Wernet {\it et al.} \cite{Wernet04} has 
recently introduced an extremely interesting new component into the study
of liquid water, by relating a pre-edge feature in the XAS spectra with 
broken HBs.
  The authors propose to determine connectivity by looking at an
electronic-structure signature of the HBs.
  Their conclusion is daring: the average coordination in liquid water would
be $\sim$2 instead of the previously accepted value slightly under 4, 
displaying a filamentous picture, instead of the distorted, partly broken 
and fluctuating tetrahedral network described in so many papers before 
\cite{waterbook,waterexp1,waterexp2,mvfs04}.

\begin{figure}[ht]
  \includegraphics*[width=7.5 cm, height=7.0 cm]{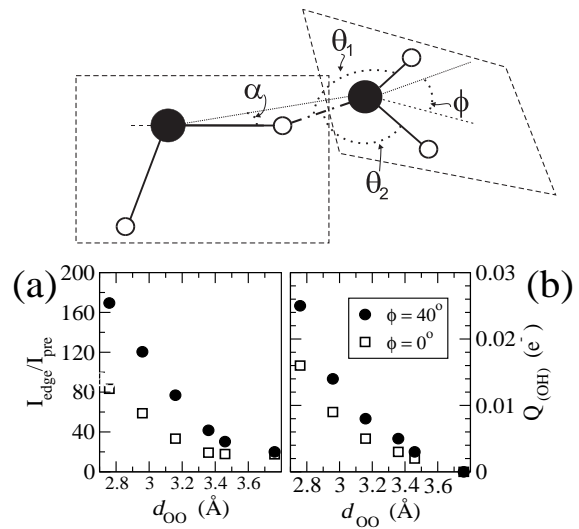}
  \caption{XAS peak intensity ratio $I_{\rm edge}/I_{\rm pre}$ (a), and
           bond order $Q_{\rm OH}$ (b) versus intermolecular distance for
	   two values of the flap, $\phi=40^{\rm o}$ and $0^{\rm o}$, for
	   the HB between an acceptor molecule and a clean (001) ice surface.}
  \label{Inten}
\end{figure}

  Is it true? This would be the wrong question to ask. 
  The kind of network depends on the definition of the hydrogen bond, 
furthermore, on deciding whether two given molecules in a given 
configuration are bonded or not. 
  There is no direct physical HB observable and there is arbitrariness in 
the choice of what is actually measured.
  Instead, we address the question of how relevant is the newly proposed 
network image for the description of the liquid in the sense of the insights 
it offers.
  The conventional criterion \cite{Luzar00} for HB is based on geometric 
considerations: an oxygen-oxygen distance within the first peak of the O-O RDF,
and an upper critical bend angle $\alpha$ (see Fig.~\ref{Inten}).
  This ``geometric" definition is based on total energy considerations
in contrast to the newly proposed ``electronic" one.
  In this paper we explore the adequacy of the newly proposed probe, 
including its time scale, virtually instantaneous as compared with atomic
motions.


  Electronic structure calculations have been performed based on 
density-functional theory (DFT), within the generalized-gradient 
approximation (BLYP) \cite{blyp1,blyp2}. 
  The {\sc Siesta} method is used \cite{Ordejon96,SiestaJPCM} with a basis 
set of atomic orbitals at the double-$\zeta$ polarized level \cite{edubases}.
  For liquid water at ambient conditions, ab initio molecular-dynamics
(AIMD) simulations have been performed in the microcanonical ensemble,
based on the DFT forces and the Born-Oppenheimer approximation.
  Further details are found in Ref.~\onlinecite{mvfs04}.
  XAS spectra have been calculated for selected configurations (see below).
  The pseudo-atomic orbitals in the basis set have been PAW-transformed 
\cite{Bloechl} into all-electron atomic orbitals for calculating matrix 
elements.
  For our basis set, neglecting the very small inter-molecular matrix 
elements was found to give an adequate approximation for the purposes of 
this paper.
  The strong excitonic effect introduced by the attraction between the core 
hole and the excited electron is estimated in the $Z+1$ approximation
\cite{Cavalleri05}.
  

  Notwithstanding the importance of the XAS experimental data for this
and other purposes, the probe provides a rather indirect
measure of the electronics of the hydrogen bond, not least because of
the mentioned excitonic effect.
  This and other difficulties\cite{Saykally04, Hetenyi04} (broadening, alignments) make it very 
difficult to obtain quantitative comparisons for the liquid phase.
  We find it more useful for our purposes to use a ground-state probe that 
we validate against XAS data in cleaner systems.
  This validation is two-sided. 
  On one hand we test our chosen probe, on the other, we test the extent
to which the XAS probe reflects the properties of the electronic ground
state.

  A very natural choice for describing electronic bonding within our method
is bond order. 
  In its simplest definition \cite{Mulliken}, the bond order between two 
atoms, 1 and 2, is $Q_{1,2}=\sum_{\mu}^1 \sum_{\nu}^2 
\rho^{\mu \nu} S_{\nu \mu}$, where $\mu$ ($\nu$) sums over the basis functions 
associated to atom 1 (2), and $\rho$ and $S$ are the density and 
overlap matrices, respectively.
  Bond orders depend on the choice of basis, and their arbitrariness has been 
described at length (see \cite{Baerends} and refs. therein).
  It is relative changes in that value what we use in this work, and
these are shown below to be meaningful enough to support its conclusions.
  Mulliken's bond orders are compared with Mayer's \cite{Mayer83}, from which 
the same conclusions are drawn.
  We believe that any other electronic-structure signature of the
bonding \cite{Baerends} would reflect the same physics.


  There has been a controversy \cite{Isaacs99,Ghanty00,Romero01,NilssonJCP05}
on the covalence of the HB and on its bonding or antibonding 
character, which would seem to affect our choice of electronic probe.
  The electronic characteristics of the HB were nicely illustrated
with maximally localized Wannier functions \cite{Romero01}.
  They can also be described in terms of an intramolecular
polarization (rehybridization within the molecules)
and intermolecular polarization or charge transfer (admixture
of orbitals of different molecules), if using the
language of single-molecule orbitals \cite{Hunt03}, or of atomic
orbitals, in either mono- or multi-determinantal wave-functions 
\cite{Ghanty00}.
  The physics of the interactions behind the HB is, however,
quite clear if one avoids the semantic problems that have been partly
behind that controversy. 
  In a typical HB there is an important electrostatic attraction that 
dominates the energetics \cite{Romero01}.
  In addition, there is a deformation of the electronic cloud
around O's accepting lone pair towards the donating H, in response
to the field generated by the latter.
  There are other effects (e.g. quantum fluctuations of the protons)
that are energetically less significant in principle, but could still
be important for liquid water. 
  They are beyond the scope of this paper.
  The mentioned polarization of the lone pair happens at the expense 
of a slight contraction of the electron cloud involved in the O-H bond 
within the donor molecule, due to Pauli exclusion \cite{Romero01}.
  This last deformation is the one behind the antibonding
character of the HB discussed in the literature \cite{Ghanty00}.
  However, the original deformation of the O's lone-pair towards the proton
remains clearly bonding.
  In this study we thus concentrate on $Q_{\rm OH}$, between the 
donor H and the acceptor O.


  In order to compare XAS and $Q_{\rm OH}$ we have calculated XAS 
spectra for a four-layer (001) ice slab, with the surface molecules 
exposing non-donated protons (dangling HBs).
  Repeated slabs are separated by 15 \AA\ of vacuum. 
  An extra water molecule has been placed on top of one of
the surface molecules, and its position has been varied. 
  Figure~\ref{Inten} shows the XAS calculated for the donor surface
molecule and $Q_{\rm OH}$ for that HB as a function of HB
O-H distance for two different values of the flap angle, $\phi$
(see Fig.~\ref{Inten}).
  $\phi\approx 40^{\rm o}$ corresponds to an ideal tetrahedral arrangement 
as in ice.
  A zero flap angle gives a rather unfavorable situation, since 
the proton faces midway between both lone pairs of O (in the nodal plane of 
the acceptor's HOMO orbital, of $C_{2v}$'s $B_1$ character), which partly 
inhibits the electronic deformation.

  We focus on two distinctive features of the XAS spectrum directly related 
to the pre-edge observed in the experiments: 
(i) the relative intensity of the peaks related to edge and pre-edge
features, and
(ii) the energy difference between both peaks $\Delta E$.
  As discussed by Cavalleri \ea \cite{Cavalleri05}, the spectral intensity
is taken from the $Z$ approximation (initial-state), while $\Delta E$
is obtained from the $Z+1$ approximation, given its sensitivity to 
final-state effects.
  Fig.~\ref{Inten} shows that $\phi=0^{\rm o}$ produces a pre-edge
twice as large as $\phi=40^{\rm o}$ for the same $d_{\rm OO}$.
  This remarkable effect is closely reproduced by $Q_{\rm OH}$, as well as 
the distance dependence.
  $\Delta E$ increases with distance in a similar manner (not shown), also 
well replicated by the bond order (the effect of the flap angle is less 
noticeable in this case).
  A detailed study of the dependence of $Q_{\rm OH}$ on intermolecular 
geometry in a water pair can be found in Ref.~\onlinecite{mythesis}.
  It is important to note that both magnitudes (XAS and $Q_{\rm OH}$) 
agree in not displaying any obvious feature (discontinuity, zero, 
minimum) that would define a natural threshold for HB breaking.
  We will thus refrain from establishing an arbitrary criterion for the
moment and explore what can be learnt independent of it.

\begin{figure}[ht]
  \includegraphics*[width=7.5 cm, height=7.0 cm]{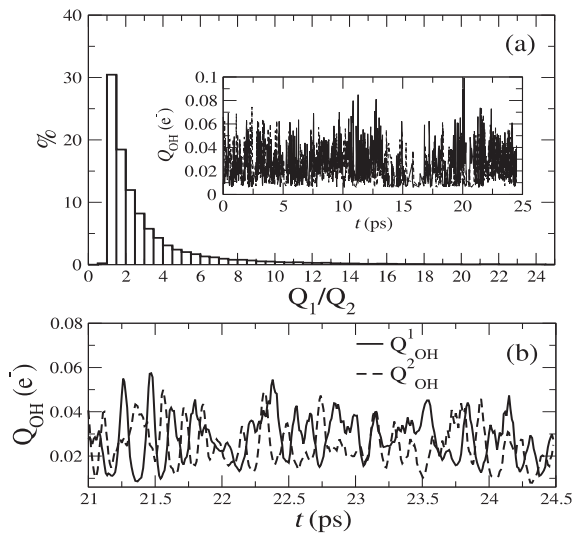}
  \caption{(a) Distribution of strong to weak bond-order ratio in the 
  liquid during a 25 ps AIMD run. (b) Evolution of the donating 
  $Q_{\rm OH}$'s for a given molecule [zoom of the inset].}
  \label{Q1/Q2}
\end{figure}

  The $Q_{\rm OH}$'s are then calculated in an AIMD simulation of liquid water 
for all water pairs within a first coordination shell, as defined by the first 
peak in the O-O RDF. 
  The first point that becomes apparent is that every water molecule is mainly 
donating one strong HB, while the second bond order is 2.2 times weaker in 
average, partly supporting the one-dimensional network picture proposed in
Ref.~\onlinecite{Wernet04}.
  The asymmetry is, however, not extreme, as can be seen in the 
distribution of strong to weak bond-order ratio in Fig.~\ref{Q1/Q2}.

  Fig.~\ref{Q1/Q2}(b) shows a fragment of the time evolution of the 
$Q_{\rm OH}$ on the two donated HBs for a representative molecule (the 
complete trajectory is shown in the inset).
  The figure shows a clear oscillation with a period of $\sim$~170 fs.
  It would correspond to the intermolecular $r$ and $t$ vibrations in ice 
(hindered rotation and translation) \cite{Silvestrelli99jcp}. 
  An intermolecular O-H under-damped oscillation with a period of
$\sim$170 fs has been indeed directly observed in liquid water \cite{Fecko03} 
using ultrafast infrared spectroscopy.
  The figure shows that the oscillation in $Q_{\rm OH}$ is more pronounced
than what expected from Fig.~\ref{Q1/Q2} (a), with lower values 5 times
smaller than the higher ones in average. 
  It shows that an anti-phase vibration of both donating HBs only accounts
for part of the oscillation, the remaining part coming from the in-phase 
vibration that would weaken (strengthen) both donated HBs simultaneously.
  The figure also shows that a single HB survives many such extreme cycles
before breaking (the average life-time for a HB is a few ps).
\cite{mythesis,Luzar00}.

  HBs with low enough values of instantaneous $Q_{\rm OH}$ are certainly 
contributing to the measured XAS pre-edge.
  The XAS probe is thus reflecting a very pronounced electronic effect, 
a swinging of the electron cloud of the lone pair following the ``flapping" 
and stretching of HBs, which appears to XAS as if many HBs were broken.
  Considering the anti-phase component, that behavior would give 
rise to a pulsating 1D filament-like network image, relevant for 
{\it electron} dynamics.
  It is, however, irrelevant to the liquid dynamics, since it is apparent 
in Fig~\ref{Q1/Q2} that the molecules in a HB are still effectively 
bound even when the electron deformation is very small.
  The inclusion of quantum fluctuations to the nuclear dynamics could even
exaggerate the beating effect, since quantum and thermal fluctuations are 
comparable in scale (the zero point motion of a 170 fs oscillation is 12 meV).

\begin{figure}[ht]
  \includegraphics*[width=7.5 cm]{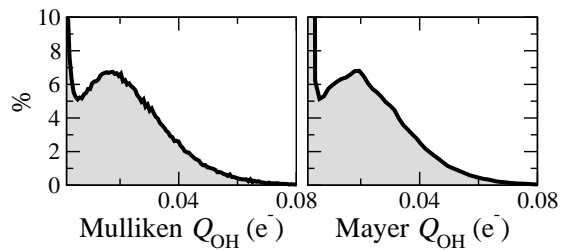}
  \caption{Distribution of bond orders in a 25 ps AIMD liquid simulation, 
  using Mulliken's \cite{Mulliken} (a) and Mayer's \cite{Mayer83} (b).}
  \label{Histogram}
\end{figure}

  Figure~\ref{Histogram} shows the distribution of $Q_{\rm OH}$ for all
the water pairs in the liquid with $r_{\rm OO} < 3.5$ \AA, for two different 
definitions of bond order.
  In both cases there is a clear minimum at small values of the
bond order, with $\sim$25~\% HBs below that threshold.
  It offers a natural criterion (albeit still arbitrary) for determining
the presence of a HB, namely, $Q_{\rm OH} > Q_{\rm OH}^{\rm min}$
\cite{frasecilla}.
  Interestingly, the width of the distribution for the HBs
($2 \Delta Q_{\rm OH} = 0.030$, $\bar Q_{\rm OH} = 0.023$) is comparable 
to that for a single oscillating HB through time (the one in 
Fig.~\ref{Q1/Q2}: $2 \Delta Q_{\rm OH} = 0.026$, $\bar Q_{\rm OH} = 0.020$).
  This means that the very different HB strengths are not so much due
to different configurations as the liquid flows (on the time scale of several 
ps, the HB lifetime) but remarkably related to these 170 fs vibrations.

\begin{figure}[ht]
  \includegraphics*[width=7.5 cm]{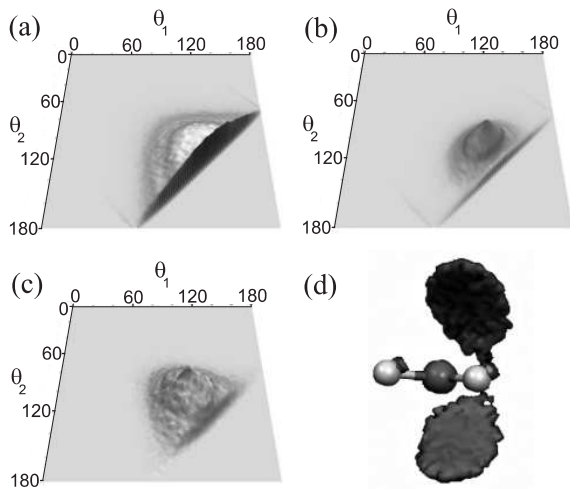}
  \caption{Distribution of $\theta_1$ and $\theta_2$ angles (as in 
           Fig.~\ref{Inten}) in liquid water, for (a) SPCE \cite{SPC/E} and 
           (b) TIP5P \cite{tip5p} force fields, and (c) AIMD. Darker
           regions indicate higher values. The $\theta_1=\theta_2$
           diagonal represents pure flap. Moving normal to that diagonal 
           represents twist. The maximum value along
           the diagonal is for zero flap (abrupt cut for rigid molecules).
           (d) Isosurface of the AIMD distribution of donating H's around a 
	   molecule.}
    \label{flapComp}
\end{figure}

  After monitoring the electronic deformation as 
signature for connectivity, we finish this study by assessing the 
effect of such deformation on the liquid structure itself.
  Having observed how the electronic cloud deforms in response 
to stretch and flap, it should be expected that the energetics would be 
affected by the flap, and so would the configurations visited in AIMD 
trajectories.
  In Fig.~\ref{flapComp} the distribution of flap angles obtained from 
AIMD and two classical non polarizable models are compared
\cite{classicalSim}.
  The angles $\theta_1$ and $\theta_2$ (as defined in Fig.~\ref{Inten}) within 
the first coordination shell are used to characterize both flap and twist.
  The distributions show clear differences.
  The main difference between SPCE and TIP5P is the fact that the latter
puts negative charges around the center of the electron lone pairs,
which induces a more realistic flap response.
  It is, however, exaggerated, since the polarization is static.
  AIMD distributions reflect both the preference for the tetrahedral
geometry and the flexibility given by the dynamical response of the
electron cloud.
%
%
  It has been argued \cite{Wernet04,Cavalleri05,Cavalleri04,Nilsson05}
 that, in spite of these differences, AIMD lines up with any force field 
so far to produce qualitatively wrong configurational sampling of liquid 
water.
  It is certainly true that there are clear shortcomings in our scheme
(BLYP approximations for exchange and correlation, neglect of protonic 
quantum effects) as well as in others.
  What we propose here, however, is a highly plausible explanation of the 
experimental results that does not imply the paradigm shift proposed by 
these authors.

  In summary, using our electronic probe for HB connectivity we observe
what could be described as 1D filamentous structures, but they are
pulsating in a 170 fs period, the geometric connectivity surviving intact 
for many periods and thus many reshapings of the filaments.
  Even if this image of pulsating filaments is not relevant for the 
description of the liquid, it is likely that questions addressing its
electronic structure could benefit from it.

\acknowledgments

  We thank J. M. Soler and X. Blase for useful discussions and
M. Dawber for help with the figures.
  EA acknowledges the hospitality at the Donostia International Physics
Centre.
  We acknowledge financial support from the British Engineering and
Physical Sciences Research Council, the Natural Environment Research
Council through the $e$Minerals project, the Cambridge European Trust, 
the R\'egion Rh\^one-Alpes, and the Comunidad Aut\'onoma de Madrid.
  The calculations were performed in the Cambridge Cranfield
High Performance Computing Facility.


\begin{thebibliography}{10}
\bibitem{frasecilla}
If one does use the $Q_{\rm OH}^{\rm min}$ criterion, HBs do not ``break" while
  oscillating. In fact, in $Q_{\rm OH}${$\sim$}$Q_{\rm OH}^{\rm min}$ cases,
  the XAS curve in Fig.~\ref{Inten} saturates to the broken HB value,
  supporting this criterion. For $\phi$=40$^{\rm o}$ this happens at $d_{\rm
  OO}${$\sim$}3.8 \AA\ (instead of the 3.3 \AA\ of Ref.~\onlinecite{Wernet04}),
  which would give less broken HBs (average coordination of 3.5 for our AIMD
  run) even if the configurations were the same.
\end{thebibliography}
\end{document}